\documentclass[]{aa}
\usepackage{keyval}

\usepackage[english]{babel}
\usepackage{graphicx}
\usepackage{txfonts}
\usepackage{amssymb}
\usepackage{enumitem}
\usepackage{graphicx}
\usepackage{amsfonts}
\usepackage{amsmath}
\usepackage{mathrsfs}
\usepackage{xcolor}
\usepackage{subcaption}
\usepackage{lscape}
\usepackage{placeins}
\usepackage{enumitem}
\usepackage{hyperref}
\usepackage{caption}

\begin{document}

\title{Evidence for a nova super-remnant surrounding the recurrent \\ nova V745\,Scorpii using eROSITA}

\author{Michael W. Healy-Kalesh\inst{1,2}\fnmsep\thanks{Corresponding author: M.W.HealyKalesh@ljmu.ac.uk. Honorary Visiting Research Fellow at Liverpool John Moores University.}
        \and Sara Saeedi\inst{1}\fnmsep\thanks{E-mail: sara.saeedi@fau.de}
        \and Manami Sasaki\inst{1}
        \and Matthew J. Darnley\inst{2}
        \and Michael M. Shara\inst{3}
        \and \\ Werner Becker\inst{4,5}
        \and Martin G. F. Mayer\inst{1}
        \and Federico Zangrandi\inst{1}
        }
\institute{Dr. Karl Remeis Observatory, Erlangen Centre for Astroparticle Physics, Friedrich-Alexander-Universität Erlangen-Nürnberg, \\ Sternwartstraße 7, 96049 Bamberg, Germany
\and Astrophysics Research Institute, Liverpool John Moores University, Liverpool, L3 5RF, UK
\and Department of Astrophysics, American Museum of Natural History, CPW \& 79th street, New York, NY 10024-5192, USA
\and Max-Planck-Institut für extraterrestrische Physik, Gießenbachstraße 1, 85748 Garching, Germany
\and Max-Planck-Institut für Radioastronomie, Auf dem Hügel 69, 53121 Bonn, Germany}

\date{Received 6 September 2026 / Accepted 23 October 2026}

\abstract{Nova super-remnants (NSRs) are vast shell-like structures encompassing recurrent novae (RNe), created through the excavation of large amounts of local interstellar medium by repeated nova eruptions. Simulations predict that such NSRs should surround all novae. Indeed, NSRs have now been discovered around the Galactic RNe KT Eridani, RS Ophiuchi and T Coronae Borealis, and the Large Magellanic Cloud RN LMCN 1971-08a, to place alongside the prototypical NSR in M\,31, associated with the extreme RN, M\,31N 2008-12a. Highly energetic novae ejecta impacting the inner boundary of the NSR shell alongside inter-ejecta shocks are predicted to generate high levels of X-ray emission in the form of an extended shell-like structure. A search for X-ray emission from a present NSR surrounding the Galactic RN, V745\,Scorpii, using observations from the extended Roentgen Survey with an Imaging Telescope Array (eROSITA) on board {\it Spektrum-Roentgen-Gamma} ({\it Spektr-RG}, {\it SRG}) led to the identification of a shell coincident with the nova. Based on (i) the structure being aligned with the location of the central nova; (ii) its size being consistent with the sizes of other known NSRs; (iii) the similarity between the emission radial profiles of the X-ray structure and the confirmed NSR around M31N 2008-12a; (vi) the presence of a corresponding infrared cavity; (v) the resemblance of the radial profile of the synthetic X-ray ring derived from a modelled NSR structure with the radial profile of the observed ring; and (vi) the lack of other sources capable of creating such a large structure, we propose this X-ray structure to be another Galactic NSR to be uncovered, and the first NSR to be captured through its X-ray emission.}


\authorrunning{Healy-Kalesh et al.}
\titlerunning{Evidence for an NSR around V745 Sco using eROSITA}

\maketitle

\section{Introduction}\label{Introduction}
Classical novae (CNe) and recurrent novae (RNe) are luminous transient events arising from the surfaces of accreting white dwarfs (WD; \citealt{1995cvs..book.....W}, and see e.g. \citealt{2020AdSpR..66.1147D}). A thermonuclear runaway triggers at the base of the accreted material once a critical mass is reached \citep{1981ApJ...243..926S}, resulting in a portion of the accumulated material being ejected in the form of a nova eruption \citep{1972ApJ...176..169S,1978A&A....62..339P}. Observationally, CNe have been observed in eruption once and RNe have exhibited more than one eruption in the last hundred years \citep{2010ApJS..187..275S,2021gacv.workE..44D}. Yet theoretically, all novae are recurrent and can have periods between eruptions ranging from $10^{-1}$ yr \citep{2016ApJ...819..168H} to $10^6$ yr \citep{2020NatAs...4..886H}. Up to ${\sim}10^{-4} \ \text{M}_{\odot}$ of the accreted material is expelled during a nova eruption and can reach velocities of up to 10,000 km\,s$^{-1}$ \citep{2001IAUS..205..260O}. Such ejecta consequently expands to form a shell around the nova (\citealt{1985ApJ...292...90C,1995MNRAS.276..353S,1998MNRAS.300..221G,2000AJ....120.2007D} and see \citealt{2025MNRAS.539..246S} for a extensive catalogue of known optical nova shells). Though only ${\sim}10\%$ of Galactic novae have observed nova remnants, these shells can fade within decades \citep{2020A&A...641A.122T} so it is probable that all novae host a nova shell and currently only the brightest examples are detectable. Matter retained on the surface after the nova eruption can grow the WD \citep{2005ApJ...623..398Y,2016ApJ...819..168H} towards the \cite{1931ApJ....74...81C} mass and result in a type Ia supernova \citep{1973ApJ...186.1007W}.

Nova super-remnants (NSRs) are greatly-extended shells created as a result of the continual sweeping up of up to thousands of solar masses \citep{2019Natur.565..460D,2023MNRAS.521.3004H} of local interstellar medium (ISM) by many prior RN eruptions \citep{2019Natur.565..460D,2023MNRAS.521.3004H,2024gacv.confE..35H}. Unlike nova shells observed around CNe (e.g. HR Del: \citealt{2003MNRAS.344.1219H} and V4362 Sgr: \citealt{2020MNRAS.499.2959H}), RNe (e.g. RS Oph: \citealt{2006Natur.442..279O,2007ApJ...665L..63B,2009ApJ...703.1955R} and T Pyx: \citealt{1997AJ....114..258S,2015ApJ...805..148S}) and dwarf novae (e.g. Z Cam: \citealt{2007Natur.446..159S,2024MNRAS.529..212S} and AT Cnc: \citealt{2012ApJ...758..121S}) which are all less than a parsec across, NSRs are defined as such on account of extending tens of parsecs in diameter \citep[see][for a review of NSRs]{2024gacv.confE..35H}: in particular, the most rapidly recurring nova, M31N 2008-12a, hosts the prototypical NSR with dimensions of 90 $\times$ 134 pc \citep{2019Natur.565..460D}. Crucially, all RNe should host an NSR \citep{2020AdSpR..66.1147D,2023MNRAS.521.3004H}.

Though a narrowband optical search (using H$\alpha$ and $[{\mathrm{S}\,\textsc{ii}}]$ imaging) in M31 and the Large Magellanic Cloud did not recover any further examples in these galaxies, this was likely due to their intrinsic faintness within the data utilised \citep{2024MNRAS.528.3531H}, later corroborated using deeper narrowband imaging around the second most rapidly recurring nova M31N 2017-01e \citep{2026RNAAS..10...60D}.

This apparent dearth has been emphatically addressed within the Galaxy by the Condor Array Telescope \citep[Condor;][]{2023PASP..135a5002L}, which can image to very faint surface brightnesses and therefore recover more of the fainter NSRs. Within a short period, Condor has demonstrated that the eleventh Galactic RN, KT Eridani, owns a 50\,pc NSR \citep{2024MNRAS.529..224S,2024MNRAS.529..236H}; the Galactic RN, T Coronae Borealis hosts a $\sim$30\,pc diameter NSR \citep{2024ApJ...977L..48S} and the Galactic symbiotic RN, RS Ophiuchi, is surrounded by an NSR 70\,pc across \citep{2025AJ....170...56S}. Most recently, the first NSR in the LMC has been found within Magellanic Cloud Emission Line Survey \citep[MCELS;][]{1999IAUS..190...28S} and DeMCELS survey \citep{2024ApJ...974...70P} H$\alpha$ and $[{\mathrm{S}\,\textsc{ii}}]$ images around the RN LMCN 1971-08a and is the largest NSR to date, at ${\sim}200$ pc in diameter \citep{2025A&A...702L...9H}.

The continuous stream of NSR discoveries will undoubtedly continue with Condor, supported by possible detections with other optical data, but the search is not constrained to the optical regime. In fact, far-IR data was able to reveal the potential cavity component of the NSR around RS Oph \citep{2024MNRAS.529L.175H} and T Pyx \citep{2026RNAAS..10..277H}. Additionally, the highly energetic ejecta impacting the high-density inner edge of the NSR shell, in combination with inter-ejecta shocks, will provide a substantial amount of X-ray emission \citep{2021PhDT........26H,2023MNRAS.521.3004H,2024MNRAS.529..236H}, such that the NSR may manifest as an extended X-ray structure.

V745 Scorpii is one of the eleven known Galactic RNe \citep{2021gacv.workE..44D,2024MNRAS.529..224S} and has a recurrence period of $\sim$25 years, having exhibited nova eruptions in 1897 \citep{2024RNAAS...8...93S}, 1937 \citep{1958AnLei..21..217P}, 1989 \citep{1989IAUC.4820....1L} and 2014 \citep{2014AAN...496....1W}, and potentially during 1917 and 1963 \citep{2021gacv.workE..44D,2024RNAAS...8...93S}. The nova system consists of a $>$1.3 $\text{M}_{\odot}$ WD \citep{2015MNRAS.454.3108P} and a M6 III \citep{1989Msngr..58...34D} or M4 III \citep{1993AJ....105..320H} red giant companion and is estimated to be located at a distance of $7.8 \pm 1.8$ kpc \citep{2010ApJS..187..275S}, though recently \cite{2024MNRAS.534.1227M} state a revised estimated distance of $8.2^{+1.2}_{-1.0}$ kpc. During its 2014 eruption, the nova ejected $10^{-7} \ \text{M}_{\odot}$ \citep{2014ApJ...785L..11B,2015MNRAS.454.3108P,2016ApJ...825...95D} of material at velocities of ${\sim}4000$ km s$^{-1}$ \citep{2014ATel.5871....1A,2014ApJ...785L..11B} and declined from peak by three optical magnitudes in 9 days \citep{2014ApJ...785L..11B}. Over its multiple eruptions, V745 Sco has been observed across optical \citep{1990MNRAS.246...78S,2014MNRAS.443..784M}, near-infrared \citep{1993AJ....105..320H,2014ApJ...785L..11B,2023ApJ...954L..16B}, radio \citep{2016MNRAS.456L..49K,2024MNRAS.534.1227M}, UV \citep{2015MNRAS.454.3108P} and X-ray \citep{2015MNRAS.448L..35O,2015MNRAS.454.3108P,2016ApJ...825...95D} wavelengths as well as being one of the collection of novae to have been detected in $\gamma$-rays \citep{2014ATel.5879....1C,2018A&A...609A.120F}. Furthermore, with a WD growing in mass, V745 Sco is proposed to be a SN\,Ia progenitor \citep[see e.g.][]{2016ApJ...825...95D}.
\begin{figure}
\centering
\includegraphics[width=\columnwidth]{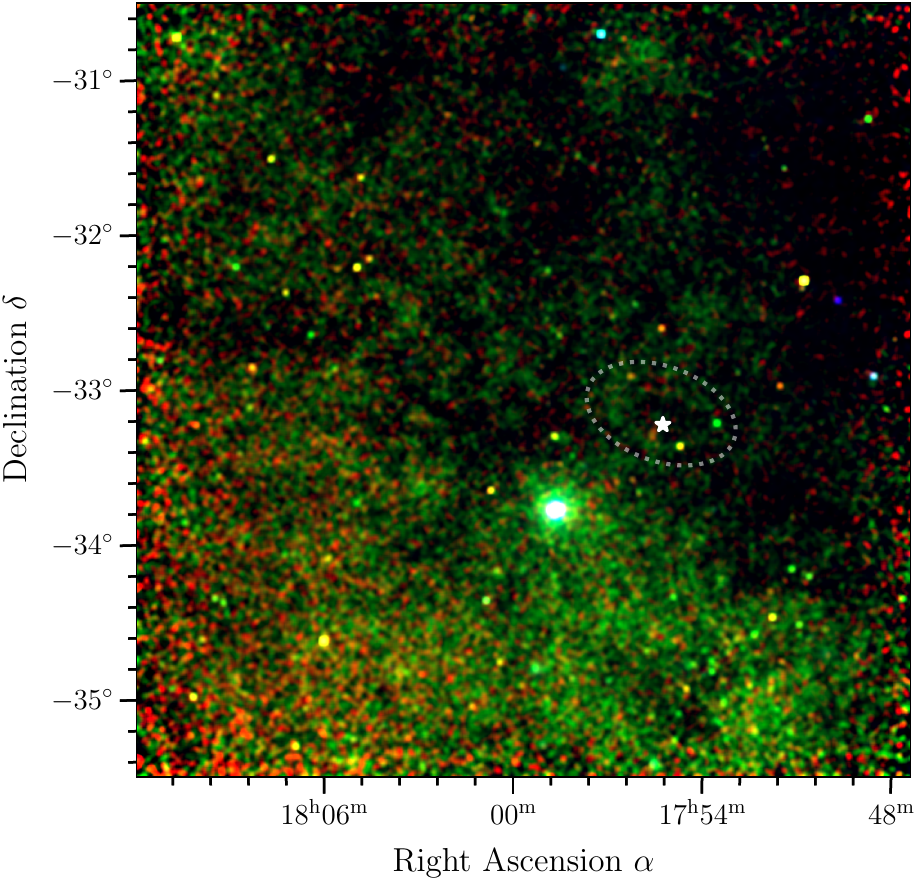}
\caption{RGB representation of eROSITA observation with a $5^{\circ} \times 5^{\circ}$ field of view covering the location of V745 Scorpii. Red, green, blue colours represent the energy bands of $0.2-0.5$\,keV, $0.5-1.0$\,keV, and $1.0-2.3$\,keV, respectively. The white star indicates the position of V745 Scorpii. A candidate NSR is visible at a coincident location to the central nova as a ring-like structure.}
\label{V745 Sco NSR eROSITA}
\end{figure}

Here we present evidence for a candidate nova super-remnant surrounding V745 Scorpii in X-ray imaging; the first NSR discovered within the X-ray regime. The paper is organised as follows: in Section~\ref{The search for an NSR surrounding V745 Scorpii}, we outline the data from which the discovery was made and derive X-ray flux upper limits of the candidate NSR, before quantifying in Section~\ref{Annular photometry of the nova super-remnant} the spatial distribution and luminosity of the emission from the structure. We follow with analyses of the NSR, starting in Section~\ref{Cavity structure in infrared data} by detailing the detection of a potentially related infrared cavity component and, in Section~\ref{Modelling of V745 Scorpii NSR}, outlining simple modelling of the shell to estimate NSR characteristics, assuming the X-ray ring and infrared cavity highlight different components of the same structure. In Section~\ref{Determining association between X-ray structure and V745 Scorpii} we then establish the likelihood of the X-ray shell being an NSR component by testing a coincidental alignment argument. We explore other origins of the X-ray shell in Section~\ref{Other origins of structure} before concluding in Section~\ref{Conclusions}.
\begin{figure}
\centering
\includegraphics[width=\columnwidth]{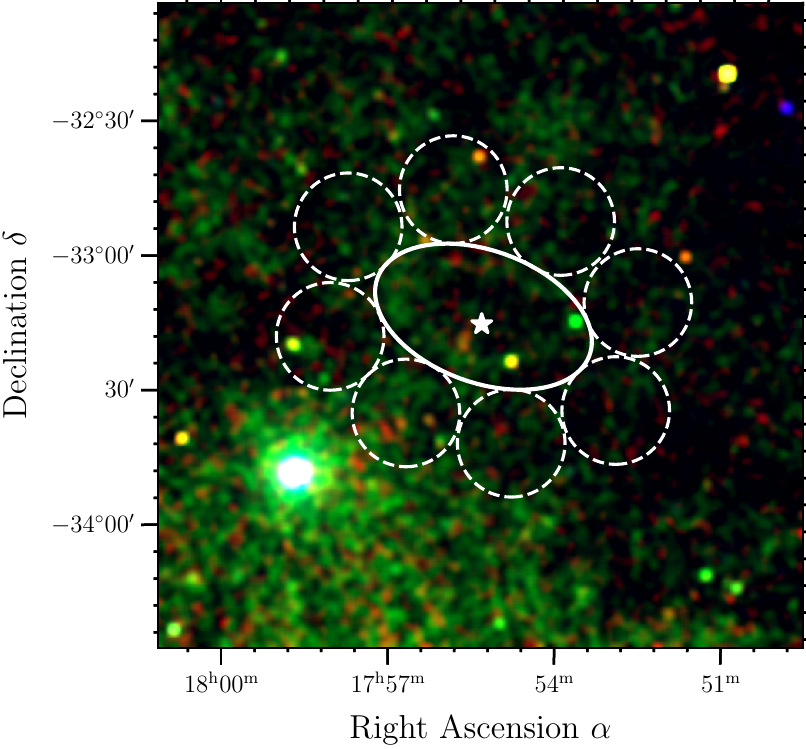}
\caption{RGB representation of eROSITA observation from Figure~\ref{V745 Sco NSR eROSITA} focussed on the X-ray ring. The solid white ellipse (${\sim}25' \times 15'$) and the dashed circles (radii of $12'$) were used as the source region and background regions, respectively, to derive a flux upper limit. The location of the nova is indicated with a white star.}
\label{V745 Sco NSR eROSITA upper limit}
\end{figure}

\section{The search for an NSR surrounding V745 Scorpii}\label{The search for an NSR surrounding V745 Scorpii}
\subsection{eROSITA data}\label{eROSITA data}
The extended Roentgen Survey with an Imaging Telescope Array \citep[eROSITA;][]{2012arXiv1209.3114M,2021A&A...647A...1P} is an X-ray telescope consisting of seven telescope modules (TM) on board the {\it Spektrum-Roentgen-Gamma} ({\it SRG}) observatory \citep{2021A&A...656A.132S}. eROSITA has an approximate $1^{\circ}$ diameter field of view covering the soft ($0.2-2.3$\,keV) and hard X-ray energy band ($2.3-8$\,keV) and since December 2019, has completed four all-sky surveys (eRASS1 $-$ eRASS4). We utilised the eROSITA Science Analysis Software System \citep[eSASS;][]{2022A&A...661A...1B,2024A&A...682A..34M} in the version \texttt{eSASSusers\_240410} for the analysis of eROSITA data in this work.

With the aim of detecting an NSR in the X-ray regime belonging to Galactic RN, V745 Scorpii, we studied the exposure-subtracted images from the eROSITA all-sky survey around this nova. These data belong to the German eROSITA consortium (eROSITA-DE) as the nova V745 Sco is located within the western Galactic hemisphere.

\subsection{Discovery image of the X-ray nova super-remnant}
In the eROSITA field of V745 Scorpii, an elliptical shell-like structure with semi-major and semi-minor axes of $\sim$19 and $\sim$10 arcminutes, respectively (and position angle of $\sim$160$^{\circ}$) was identified. As shown in Figure~\ref{V745 Sco NSR eROSITA}, the centre of the shell is coincident with the nova. For a distance to V745 Sco of $8.2^{+1.2}_{-1.0}$ kpc \citep{2024MNRAS.534.1227M}, this elliptical structure has dimensions of ${\sim}90 \times 50$ pc, therefore its size is reminiscent of other NSRs. The predominant X-ray emission from the NSR shell is in the eROSITA $0.5-1.2$ keV waveband.

\subsection{X-ray flux upper limit of the nova super-remnant}\label{X-ray flux upper limit of the nova super-remnant}
Although there is a ring seen surrounding V745 Sco, we do not see a significant X-ray excess from the region as a whole. As such, we are not able to present any spectral analysis. However, we are able to derive X-ray flux upper limits for the whole region containing the ring using two approaches: a maximum possible count rate ($\mathrm{CR}_{\mathrm{max}}$) methodology presented in \cite{2024A&A...683A.180K} and an independent estimate based on the local background fluctuations.

As in \cite{2024A&A...683A.180K}, we first used the $\mathrm{CR}_{\mathrm{max}}$ approach to ascertain a quantitative measure of any possible extended emission. We extracted photons from an elliptical source region with dimensions of ${\sim}25' \times 15'$ as shown in Figure~\ref{V745 Sco NSR eROSITA upper limit}. We determined the background level around the source region by taking the average of the counts from 8 circles with radii of $12'$ employed around the source region, also shown in Figure~\ref{V745 Sco NSR eROSITA upper limit}. We masked the two bright sources within the elliptical source region to minimise contamination.

The X-ray counts rate for the source region and background regions were found to be CR$_\mathrm{s} = 0.261 \pm 0.002 \ \mathrm{cts} \ \mathrm{s}^{-1} \ \mathrm{deg}^{-2}$ and CR$_\mathrm{b} = 0.263 \pm 0.001 \ \mathrm{cts} \ \mathrm{s}^{-1} \ \mathrm{deg}^{-2}$, respectively, with the source count rate uncertainty and background count rate uncertainty estimated as in \cite{2024A&A...683A.180K}. This analysis reveals that the source region as a whole does not show a significant excess above the surrounding background level. Therefore, as in \cite{2024A&A...683A.180K}, we calculated the maximum possible count rate using
\begin{equation*}
\mathrm{CR}_{\mathrm{max}}= (\mathrm{CR}_{\mathrm{s}}-\mathrm{CR}_{\mathrm{b}}) + 3 \sqrt{\sigma^2_{\mathrm{s}} + \sigma^2_{\mathrm{b}}}, 
\end{equation*}
where $\mathrm{CR}_{\mathrm{s}}$ is the count rate within the elliptical source region, $\mathrm{CR}_{\mathrm{b}}$ is the mean count rate within the background regions, and $\sigma_{\mathrm{s}}$ and $\sigma_{\mathrm{b}}$ are the associated uncertainties combined in quadrature, and was found to be $\mathrm{CR}_{\mathrm{max}}=6.871\times 10^{-3} \ \mathrm{cts} \ \mathrm{s}^{-1} \ \mathrm{deg}^{-2}$.

We then used data from four combined event files of eROSITA\,(eRASS1 to eRASS4) in order to derive the X-ray flux upper limit. We extracted spectra using the same source region and background regions illustrated in Figure~\ref{V745 Sco NSR eROSITA upper limit} and used all TM cameras of eROSITA except TM5 and TM7, due to the technical problem of optical light contamination in their CCDs \citep{2021A&A...647A...1P}.

The X-ray Spectral Fitting Package \citep[\texttt{XSPEC};][]{1996ASPC..101...17A} was utilised to convert the maximum possible count rate, $\mathrm{CR}_{\mathrm{max}}$, into flux. As is typically used for modelling the X-ray emitting shock-heated plasma in supernova remnants \citep{2022A&A...661A..37S,2025A&A...701A..60K}, and the most analogous structures to NSRs, we assumed an absorbed thermal plasma model for the spectrum, namely \texttt{TBabs\,*\,vapec} (with abundances fixed to solar values), with \texttt{TBabs} corresponding to X-ray absorption by the ISM \citep{2000ApJ...542..914W} and \texttt{vapec} corresponding to a collisionally ionised diﬀuse gas in thermal equilibrium, and simulated the observations using \texttt{XSPEC}. Within this model, the {$N_\text{H}$} absorption was fixed at $2.8\times10^{21}$\,cm$^{-2}$ to match the Galactic absorption in the direction of V745\,Scorpii \citep[from the HI4PI Survey;][]{2016A&A...594A.116H}. The observations were simulated using \texttt{XSPEC} by assuming this model and examining various values of plasma temperature ($k$T $= 0.2, 0.5$ and $1$ keV). The $3\sigma$ flux upper limit within the energy band $0.5 - 2$ keV, representing the sensitivity to emission averaged over the full source region, was found to be $(3.5-3.8)\times10^{-14}$\,erg\,s$^{-1}$ cm$^{-2}$, corresponding to a luminosity upper limit of ${\sim}2.9\times 10^{32}$\,erg\,s$^{-1}$ at a distance of $\sim$8.2\,kpc to the nova (see Section~\ref{Introduction}).

Secondly, as an independent estimate for the detection threshold, we calculated an upper limit using the background fluctuations measured in the local background regions. The standard deviation was determined from the distribution of background count rates and then scaled to the area of the source region. Assuming the same \texttt{XSPEC} model described above, the 3$\sigma$ flux upper limit derived from this was ${\sim}4.2\times10^{-14}$\,erg\,s$^{-1}$ cm$^{-2}$, corresponding to a luminosity upper limit of ${\sim}3.4\times 10^{32}$\,erg\,s$^{-1}$. This is consistent with the $\mathrm{CR}_{\mathrm{max}}$ upper limit.

\section{Annular photometry of the nova super-remnant}\label{Annular photometry of the nova super-remnant}
\begin{figure}
\centering
\includegraphics[width=\columnwidth]{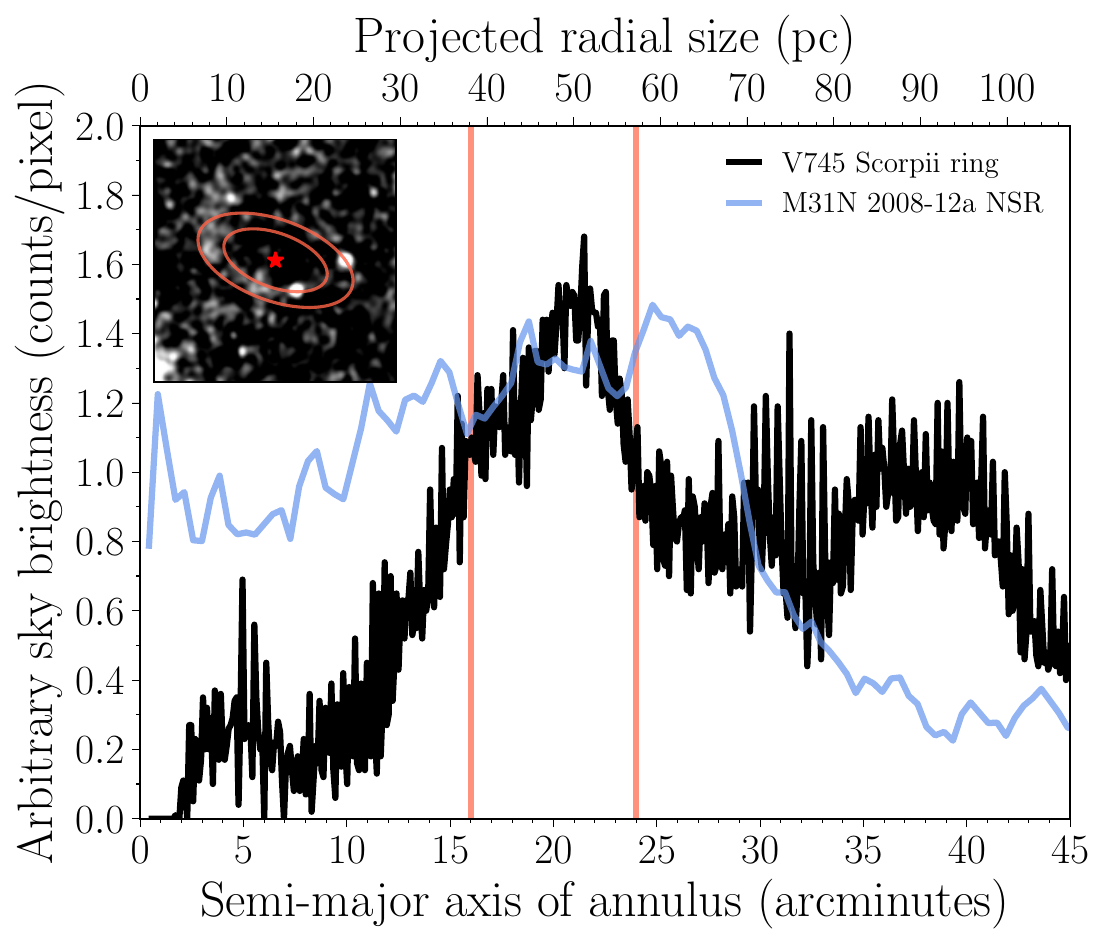}
\caption{Emission radial profile (black line) of X-ray ring around V745 Sco determined from annular photometry after continuum subtraction, smoothing and bright source masking. Vertical lines represent the location of the annuli indicated in the inset and the cross indicates the location of V745 Sco. The two bright sources in the inset do not contribute to the radial profile. The H$\alpha$ radial profile for the confirmed NSR around M31N 200-12a (blue line) is provided for comparison \citep[from][]{2024MNRAS.528.3531H}.}
\label{X-ray radial profile}
\end{figure}
Inspection of Figure~\ref{V745 Sco NSR eROSITA} shows that the shell-like structure is faint and difficult to distinguish from the background (see Section~\ref{X-ray flux upper limit of the nova super-remnant}). Therefore, to further quantify the spatial distribution of the emission and investigate the morphology of the structure, we performed annular photometry at the location of V745 Sco. The annular photometry consisted of 500 linearly spaced elliptical annuli with semi-major axes ranging from $0.32 - 48$ arcminutes ($1 - 150$ pixels), corresponding to a radial spacing of ${\sim}0.3$ pixels. All ellipses had ellipticities of 0.5 (to match the semi-major axis of 19 arcmin and semi-minor axis of 10 arcmin) and position angles of $160^{\circ}$ to match the elliptical shell in the data. Also, before the annular photometry, we applied smoothing to the data ($\sigma = 3$) and masked bright sources with pixel values $>0.0025$. 

The result of our emission radial profile analysis is shown in Figure~\ref{X-ray radial profile}. While the profile shows noticeable scatter, resulting from the large number of annuli used in the process, there is a clear peak at ${\sim}21$ arcminutes. From Figure~\ref{X-ray radial profile}, we can see that this peak in emission sits within the range of ${\sim}$16 and 24 arcminutes, consistent with the semi-major axis of the shell-like structure from inspection (19 arcminutes). The inset image of Figure~\ref{X-ray radial profile} illustrates the (narrow) annuli used for photometry with semi-major axes of 16 and 24 arcminutes. Moreover, the two bright sources in the inset image were masked prior and so do not contribute to the annular photometry.

For comparison, in Figure~\ref{X-ray radial profile} we have overlaid the H$\alpha$ surface brightness radial profile of the confirmed nova super-remnant around M31N 2008-12a from \citet{2024MNRAS.528.3531H}. This radial profile was created using the same method of annular photometry as applied to the X-ray ring around V745 Sco. Though the shell is faint, from Figure~\ref{X-ray radial profile} we see that it does stand out from the background and is reminiscent of the radial profile of the M31N 2008-12a NSR: an inner cavity region bound by an over-density and surrounded by a relatively flat ISM.

In order to quantify the X-ray luminosity of the X-ray ring, we took the count rates within each annulus and converted these into fluxes. A local background level was estimated from the background regions and then subtracted from each annulus to derive the net emission associated with the ring. By summing the annuli corresponding to the shell region (between ${\sim}16$ and 24 arcminutes) we determined the flux of the ring to be ${\sim}2\times10^{-12}$\,erg\,s$^{-1}$ cm$^{-2}$, corresponding to a luminosity of ${\sim}1.6\times 10^{34}$\,erg\,s$^{-1}$. Moreover, summing the flux contributions of all annuli within the whole region leads to a luminosity of ${\sim}2.3\times 10^{34}$\,erg\,s$^{-1}$, revealing that the X-ray ring accounts for ${\sim}70\%$ of the total emission within this region. 

These values are well above the detection thresholds derived in Section~\ref{X-ray flux upper limit of the nova super-remnant}, thereby confirming that the observed emission from the X-ray ring represents a significant excess above the background. To reiterate, the upper limits detailed in Section~\ref{X-ray flux upper limit of the nova super-remnant} refer to diffuse emission across the region, whereas the luminosities derived here from annular photometry predominantly trace the ring, which contains the majority (${\sim}70\%$) of the emission, is clearly detected above the background, and therefore is unaffected by the diffuse emission upper limits.

\section{Cavity structure in infrared data}\label{Cavity structure in infrared data}
We searched through archival multiwavelength data across numerous facilities of the surroundings of V745 Sco with fields of view wide enough to capture associated structures but did not find anything apparent. As the $23.6' \times 23.6'$ FOV of the X-ray Telescope (XRT) onboard the Neil Gehrels
{\it Swift} Observatory \citep[{\textit{Swift};}][]{2004ApJ...611.1005G} is smaller than the full extent of the X-ray ring, 100 ks of XRT data focussed on the location of V745 Sco did not capture the NSR, with only known point sources being visible. Similarly, nothing is found at the location of the X-ray structure in the Australian Square Kilometre Array Pathfinder \cite[ASKAP;][]{2007PASA...24..174J} Rapid ASKAP Continuum Survey or in neutral atomic hydrogen ($\ion{H}{i}$) data from the HI4PI survey \citep{2016A&A...594A.116H}.
\begin{figure*}
\centering
\includegraphics[width=\textwidth]{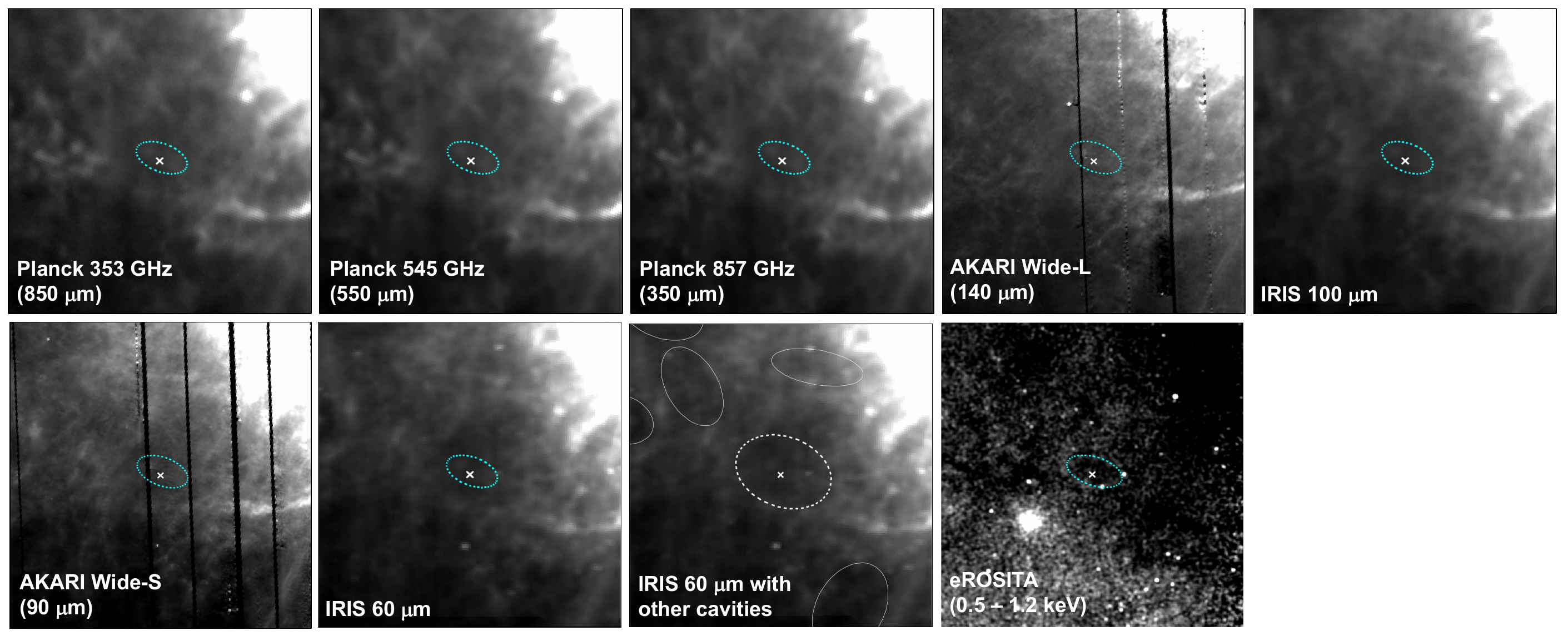}
\caption{$3.6^{\circ} \times 3.6^{\circ}$ images of the surroundings of V745 Scorpii from various infrared facilities (Planck, IRIS, AKARI) and eROSITA in the $0.5 - 1.2$ keV waveband. The boundary of the X-ray structure shown in Figure~\ref{V745 Sco NSR eROSITA} is outlined with a cyan ellipse for guidance and V745 Scorpii is indicated with a cross in each panel. The facility and wavelength is also provided in each panel. A cavity with the same position angle as the X-ray shell is evident in all infrared data and is approximately outlined in the third panel on the bottom row with a white dashed ellipse, along with other cavity-like structures indicated (see text).}
\label{IR images}
\end{figure*}

However, motivated by the NSR cavity found around RS Oph in far-infrared data \citep{2024MNRAS.529L.175H}, we also checked the location of the X-ray structure within a $3.6^{\circ} \times 3.6^{\circ}$ FOV in the 60\,$\mu$m waveband of the new generation {\it Infrared Astronomical Satellite} ({\it IRAS}) catalogue, known as IRIS \citep[Improved Reprocessing of the {\textit{IRAS} Survey;}][]{2005ApJS..157..302M} and found a larger cavity encompassing the position of the X-ray structure, shown in Figure~\ref{IR images}. A search through data from other infrared facilities, namely Planck \citep{2004AdSpR..34..491T,2011A&A...536A..21P} and AKARI \citep{2007PASJ...59S.369M,2007PASJ...59S.389K}, shows the same larger infrared cavity shadowing the location of the X-ray shell. In Figure~\ref{IR images}, we have overlaid an ellipse approximately outlining the extent of the X-ray structure (from Figure~\ref{V745 Sco NSR eROSITA}) over the infrared data for comparison.

This infrared cavity has semi-major and semi-minor axes of 35 and 25 arcminutes, respectively, which equates to a physical size of $167 \times 120$ pc at the distance of V745 Sco, and it also has a similar position angle (${\sim}160^{\circ}$) to the X-ray shell. Further, the approximate centre of the far-infrared cavity is near to the location of V745 Sco. From this apparent configuration, a simple interpretation of the two structures could be that the X-ray shell seen in eROSITA data highlights either the boundary between the inner NSR cavity and the ejecta pile-up region or a portion of the ejecta pile-up region, where we find the most extreme temperatures (see Section~\ref{Results of simulations} for more details on the different components of an NSR), and the large extended cavity revealed in the far-infrared unveils the boundary of the outer NSR shell; this interpretation is explored in Section~\ref{Modelling of V745 Scorpii NSR}. Furthermore, if this were indeed the configuration, then we would expect the optical outer shell of the NSR to be found with dimensions of ${\sim}167 \times 120$ pc.

\subsection{Probability of finding coincident extended object in infrared data}\label{Probability of finding coincident extended object in IR data}
We determined the probability of finding an extended structure in infrared data at the same position as the X-ray shell with a similar position angle. We note that the morphology of the X-ray ring and cavity appear to be consistent but we do not include this parameter in the following calculation. Given the apparent sizes of the X-ray shell and the infrared cavity, the $3.6^{\circ} \times 3.6^{\circ}$ field of view of the IRIS 60 $\mu$m, and the approximate 5 other cavity-like structures (see the bottom middle panel of Figure~\ref{IR images}) present in the data, we estimate a ${\sim}76\%$ chance of an infrared cavity within the field randomly overlapping with the X-ray shell. However, also accounting for the probability that the position angles of the X-ray shell and infrared cavity also coincide, with a conservative margin of $\pm 5^{\circ}$, we estimate the probability of finding an infrared cavity with a similar position angle overlapping spatially with the X-ray shell to be ${\sim}4.2\%$, strengthening an association between the two.

\subsection{Connection between infrared cavity and X-ray shell}\label{Connection between IR cavity and X-ray shell}
\begin{figure}
\centering
\includegraphics[width=\columnwidth]{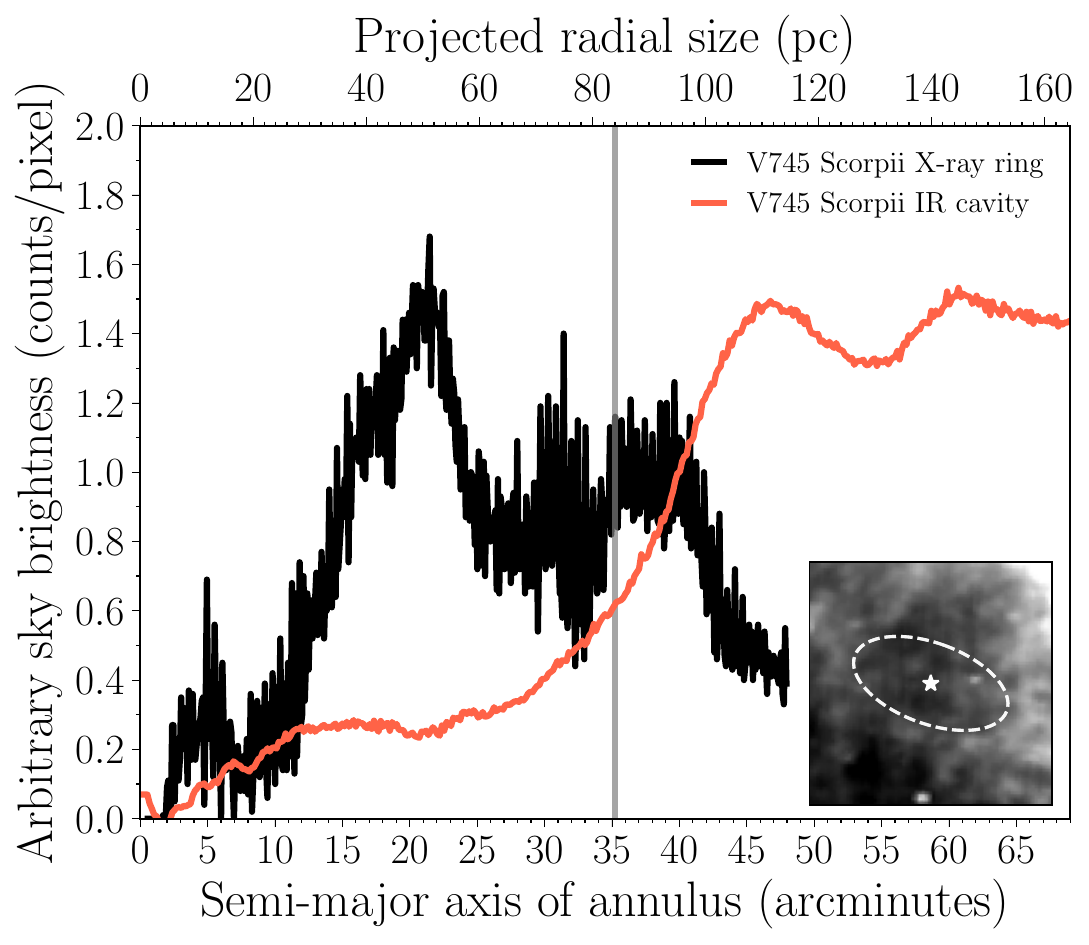}
\caption{Radial profile of the infrared cavity (red line) from annular photometry and the radial profile of the X-ray ring (black line) is provided for comparison. The grey vertical line at 84\,pc indicates the semi-major axis of the infrared cavity. Within the inset, the white star shows the location of V745 Sco and the dashed ellipse represents the ellipse with a semi-major axis of 84\,pc. The brightnesses of the two radial profiles have been scaled accordingly to focus on the radial sizes of both.}
\label{IR cavity annular photometry}
\end{figure}

As with the X-ray emission from the NSR, we performed the same annular photometry (from Section~\ref{Annular photometry of the nova super-remnant}) to the far-infrared data focussing on the cavity. The resulting radial profile is illustrated in Figure~\ref{IR cavity annular photometry}, alongside the radial profile obtained for the X-ray ring (from Figure~\ref{X-ray radial profile}).

The infrared radial profile (red line) in Figure~\ref{IR cavity annular photometry} clearly shows lower levels of emission at the centre of the cavity which gradually increases out to a peak of ${\sim}$46 arcminutes (corresponding to $\sim$100 pc), before flattening off to a relatively constant brightness beyond. This emission profile verifies the cavity-like region seen in the 60\,$\mu$m IRIS data. Within this infrared cavity sits the peak of the X-ray ring at $\sim$21 arcminutes. This configuration -- an X-ray ring encompassed by an infrared cavity -- is reminiscent of NSR modelling (presented in Section~\ref{Modelling of V745 Scorpii NSR}) that predicts a structure consisting of an X-ray shell where the highest levels of X-ray emission persist (ejecta pile-up region through to the inner edge of the NSR outer shell) surrounded by an infrared cavity (where the outer shell emits in the optical). Furthermore, an NSR with a non-spherical symmetry would allow for variations of the projected locations of the different NSR components.

\section{Modelling of a V745 Scorpii NSR}\label{Modelling of V745 Scorpii NSR}
To determine the gross properties of the NSR from its angular size in both the eROSITA X-ray and IRIS infrared data, we ran a set of hydrodynamical simulations with the \texttt{Morpheus} code \citep{2007ApJ...665..654V}, in a similar manner to previous modelling studies of NSRs \citep{2019Natur.565..460D,2023MNRAS.521.3004H,2024MNRAS.529..236H,2025A&A...702L...9H}, to evolve an NSR to the extent of the larger infrared cavity (containing the X-ray ring component).
\begin{table*}
\captionsetup{justification=raggedright, singlelinecheck=false}
\caption{Parameters for each hydrodynamical simulation of a growing NSR within surrounding ISM of different densities.}
\label{Runs}
\centering
\begin{tabular}{cccccccc}
\hline
Run \# & ISM density & Pile-up boundary & Shell inner edge & Shell outer edge & Eruptions & Elapsed time & Total KE\\
& & (pc) & (pc) & (pc) & & (yr) & (erg) \\
\hline
1 & 0.001 & 16.9 & 77.4 & 83.9 & \phantom{00}82,615 & $2.2 \times 10^{6}$ & $1.3 \times 10^{48}$ \\
2 & \phantom{0}0.01 & 9.4 & 64.6 & 84.2 & \phantom{0}158,504 & $4.1 \times 10^{6}$ & $2.5 \times 10^{48}$ \\
3 & \phantom{00}0.1 & 8.3 & 72.4 & 84.0 & \phantom{0}353,512 & $9.2 \times 10^{6}$ & $5.6 \times 10^{48}$ \\
4 & \phantom{000}1 & 6.4 & 82.4 & 83.9 & 1,009,812 & $2.6 \times 10^{7}$ & $1.6 \times 10^{49}$ \\
5 & \phantom{00}10 & 4.7 & 80.2 & 84.0 & 2,441,530 & $6.4 \times 10^{7}$ & $3.9 \times 10^{49}$ \\
\hline
\end{tabular}
\tablefoot{Columns record the simulation number, ISM density, boundary of cavity and ejecta pile-up region, boundary of inner NSR shell, boundary of outer NSR shell, number of eruptions to grow the nova super-remnant to approximately 84 parsecs (the angular semi-major axis of the outer boundary of the infrared NSR cavity around V745 Sco), the elapsed (cumulative) time of the simulation, and the total kinetic energy released. The recurrence period and ejecta velocity for all simulations is 26 years and 4000 km s$^{-1}$, respectively. Please refer to \citet{2023MNRAS.521.3004H} for more information on the defined boundaries of the NSR components (cavity, ejecta pile-up region and NSR shell).}
\end{table*}

We adopted a scenario of a static WD \citep[also see][]{2024MNRAS.529..236H,2025A&A...702L...9H} and took the recurrence period of V745 Sco to be 26 years\footnote[1]{We note that \cite{2024RNAAS...8...93S} states the recurrence period is 23 years when accounting for two likely missed eruptions in 1917 and 1963.} \citep{2021gacv.workE..44D}. As the mass loss phase, $t_{\text{ml}}$, of the nova eruption can be approximated by the $t_3$ decline time \citep{2021PhDT........26H}, we took $t_{\text{ml}}$ to be 9 days \citep{2014ApJ...785L..11B} and the total mass ejected during each eruption to be $10^{-7} \ \text{M}_{\odot}$ \citep{2015MNRAS.454.3108P,2016ApJ...825...95D}. For the nova ejecta velocity, \cite{2014ApJ...785L..11B} report the near-infrared Pa\,$\beta$ emission line FWHM to be $\sim$4000 km s$^{-1}$ and \cite{2014ATel.5871....1A} concur that all hydrogen Balmer (FWZI$\sim$8500 km s$^{-1}$), \ion{Fe}{ii}, \ion{He}{i} and \ion{O}{i} lines have similar profiles to the near-infrared emission lines. As such, we modelled the eruptions with an ejecta velocity of 4000\,km\,s$^{-1}$. 

With V745 Sco having a red giant companion \citep{1989Msngr..58...34D}, we adopted a wind loss rate of $2 \times 10^{-8} \ \text{M}_{\odot} \ \text{yr}^{-1}$ and a wind velocity of 20 km s$^{-1}$ (used to mimic the red giant companion of the Galactic RN RS Oph) as in previous NSR modelling \citep{2019Natur.565..460D,2023MNRAS.521.3004H,2024MNRAS.529..236H}. In these simulations, we assume one-dimensional spherical symmetry and choose a spatial resolution of 1000 au cell$^{-1}$ as previous studies illustrate that such a resolution is able to capture the full long-term structure of the NSR while helping with computational time \citep{2023MNRAS.521.3004H,2024MNRAS.529..236H}. \cite{2023MNRAS.521.3004H} determined that radiative cooling has a large impact on NSR growth when considering long inter-periods ($>$1\,year) and low-energy eruptions. Therefore, we adopted the cooling package \citep{1976ApJ...204..290R} provided in the \texttt{Morpheus} code as in previous studies \citep{2007ApJ...665..654V,2019Natur.565..460D,2023MNRAS.521.3004H,2024MNRAS.529..236H}.

The nova eruptions interject the red giant wind mass-loss at intervals set by the recurrence period and are represented by an instantaneous increase in the mass-loss rate and ejecta velocity to match the parameters estimated for V745 Sco. Specifically, each eruption from this modelled nova system ejects $10^{-7} \ \text{M}_{\odot}$ at a velocity of 4000 km\,s$^{-1}$ over 9 days every 26 years into a prepopulated ISM with varying densities. As the ISM density at the location of V745 Sco is not constrained, we chose to sample the density of the ISM over a range of $n=0.001$ to $n=10 \ \text{cm}^{-3}$ (we will drop the units throughout). The parameters of the set of simulations are detailed in Table~\ref{Runs}.

\subsection{Results of simulations}\label{Results of simulations}
\begin{figure}
\centering
\includegraphics[width=\columnwidth]{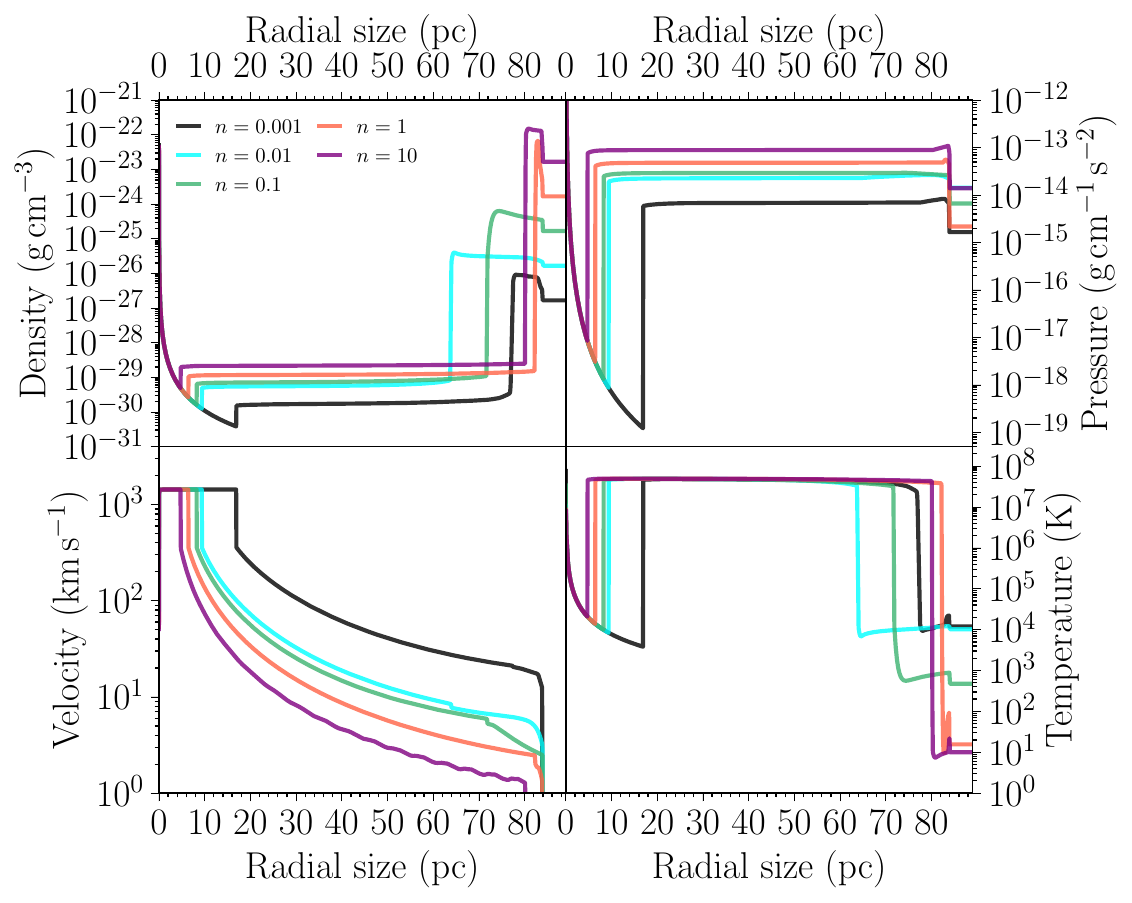}
\caption{Dynamics of the simulated NSR surrounding V745 Sco for ISM densities of $n = 0.001$, $0.01$, $0.1$, $1$ and $10$. The spatial resolution is too low to resolve the individual eruptions, but each eruption is carefully time resolved in the simulations.\protect\footnotemark[2]}
\label{V745 Sco NSR simulation}
\end{figure}

In Figure~\ref{V745 Sco NSR simulation}, we show the resulting dynamics of an NSR grown in local ISM with densities $n= 0.001, 0.01, 0.1, 1$ and $10$ to the size of the observed infrared cavity (semi-major axis of approximately 84 parsecs) in Figure~\ref{IR images}. As detailed in Table~\ref{Runs}, there is a large difference in both the amount of time and number of eruptions required to grow the NSR to 84 parsecs across the range of ISM densities. Specifically, it would have taken V745 Sco ${\sim}6.4 \times 10^7$\,yr to grow the NSR shell to the observed size in an ISM of $n = 10$ compared to ${\sim}2.2 \times 10^6$\,yr for $n=0.001$. This is approximately 30$\times$ longer (with 30$\times$ more eruptions) and results from the increasing number of eruptions required to transport the higher density surrounding material to a distance of 84 parsecs.

As shown in the top left panel of Figure~\ref{V745 Sco NSR simulation}, the thickness of the outer shell (see Table~\ref{Runs} for the location of the inner edge and outer edge) also differs dramatically across the range of ISM densities: 7.7\% for $n=0.001$, 23.3\% for $n=0.01$, 13.8\% for $n=0.1$, 1.8\% for $n=1$ and 4.5\% \ for $n=10$ due to the influence of cooling. The wider low-density shells closely match those NSRs simulated in \citet{2019Natur.565..460D} and \citet{2024MNRAS.529..236H}, in which radiative cooling does not play a substantial role because of the lower densities and shorter evolutionary times involved. Whereas the thin high-density shells seen in Run 4 and Run 5 are reminiscent of the NSR shells presented in \citet{2023MNRAS.521.3004H}, which form as a result of high levels of radiative cooling losses early in their formation suppressing the shell width over the longer eruption history (see Appendix~\ref{Radiative cooling energy losses} for details on energy losses due to radiative cooling).

In addition to the different shell thicknesses, the structure of the NSR shell varies. The inner edge of the shell is denser than the outer edge for Run 1 ($n_{\text{in}}/n_{\text{out}}\sim 2.5$), Run 4 ($n_{\text{in}}/n_{\text{out}}\sim 11$) and to a lesser extent for Run 3 ($n_{\text{in}}/n_{\text{out}}\sim 1.8$). As with the simulated NSRs presented in \citet{2023MNRAS.521.3004H} which show a similar decrease in density across the radius of the shell, this structure in Run 4 can be attributed to the outer edge depending upon a combination of the ${\sim}1\times10^6$ prior eruptions, whereas the inner edge is being shaped by the newly incoming ejecta. Though, for Run 1 and Run 3 with fewer eruptions, this similar structure likely comes from the relative ease of manoeuvring the low-density material by the ejecta into a `forming wave' shape. The density gradient of the NSR shell in Run 2 is more constant, in a similar manner to the simulated NSRs in \citet{2019Natur.565..460D} and \citet{2024MNRAS.529..236H}, and likely results from a combination of a higher ISM density compared to $n=0.001$ but a shorter cooling time compared to $n=0.1$ preventing significant compression of the shell.
\footnotetext[2]{In previous publications reporting simulations of nova super-remnants \citep{2023MNRAS.521.3004H,2024MNRAS.529..236H,2024MNRAS.529L.175H}, we note a typo on plots showing NSR dynamics involving pressure. The works were conducted using data with the cgs unit, dyne (g\,cm$^{-1}$\,s$^{-2}$), but a simple typo (g\,cm$^{-3}$\,s$^{-2}$) was made on plots. This has no impact on the data, findings nor conclusions of these publications. We have addressed this typo in this current paper.}

Within the cavity region of the NSRs (contained within the pile-up boundary; see Table~\ref{Runs} for the location of this boundary), the density of the material drops steeply from the central nova as the ejecta expand, decreasing as low as $n \simeq 2.3 \times 10^{-7}$ at the ejecta pile-up boundary for Run 1 (${\sim}4400 \times$ less dense than the density of the surrounding ISM) on account of the ejecta facing negligible resistance. Such a region of near free-expansion also explains the almost identical ejecta velocities of ${\sim}1400 \ \text{km} \ \text{s}^{-1}$ (see bottom left panel of Figure~\ref{V745 Sco NSR simulation}), the very low pressures of $1.2 \times 10^{-19} \ \text{g} \ \text{cm}^{-1} \ \text{s}^{-2} - 8.5 \times 10^{-18} \ \text{g} \ \text{cm}^{-1} \ \text{s}^{-2}$ for $n=0.001-10$, respectively (see top right panel of Figure~\ref{V745 Sco NSR simulation}) and the relatively cool temperatures of $3800-21000 \ \text{K}$ for $n=0.001-10$, respectively (see bottom right panel of Figure~\ref{V745 Sco NSR simulation}) seen in the cavity for Runs $1-5$.
\begin{figure*}
\centering
\includegraphics[width=\textwidth]{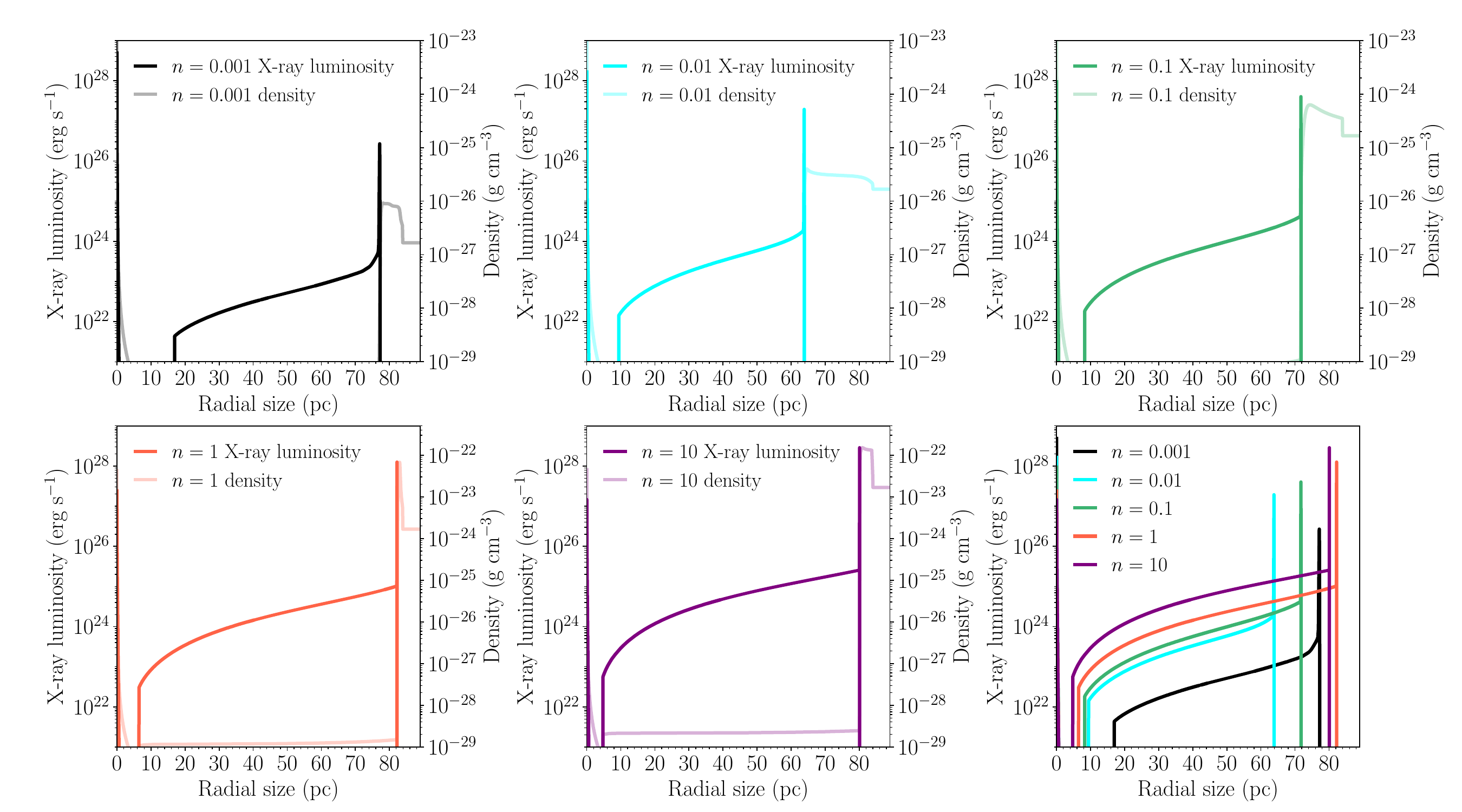}
\caption{Synthetic X-ray emission (without absorption) of the NSRs grown in Runs $1-5$ (solid lines), showing the radial distribution of cell X-ray luminosities, along with the density radial profiles (faint lines) of the NSRs (from Figure~\ref{V745 Sco NSR simulation}). The bottom right panel shows a comparison of the radial distribution of X-ray luminosities for each ISM density. For all Runs, the X-ray emission is significant close to the nova (origin) before dropping considerably in the cavity region. The higher X-ray emission then recovers within the ejecta pile-up region before substantially increasing to its peak at the inner edge of the NSR outer-shell. We note that the density ranges vary between panels for clarity.}
\label{Synthetic x-ray profile}
\end{figure*}

Beyond traversing the sparse cavity component of the NSR, the ejecta run into the pile-up region. Here, as seen in the bottom left panel of Figure~\ref{V745 Sco NSR simulation}, the ejecta velocities drop considerably for all the runs and smoothly decline through the full pile-up region until the outer shell of the NSR. There are small fluctuations in velocity for Run 5 ($n=10$) however such perturbations are ${\sim}1 \ \text{km} \ \text{s}^{-1}$ and therefore likely exist in the other runs, but are not evident due to the higher velocities present within those runs. The pressure and temperature of the matter within the pile-up region dramatically increases from the cavity pressure and temperature by ${\sim}20000 \ \times$ and ${\sim}10000 \ \times$, respectively, for all runs (see top right and bottom right panels of Figure~\ref{V745 Sco NSR simulation}) as new energetic ejecta crashing into the inner boundary and continual collisions taking place within the pile-up region leads to extreme levels of shock heating. This extreme region is where we expect the highest levels of X-ray emission from an NSR, powering the X-ray ring in the eROSITA data.

Encompassing the cavity and pile-up regions is the outer shell of the NSR. The outer shell, most apparent in the top left panel of Figure~\ref{V745 Sco NSR simulation}, is a high density region predominantly consisting of swept up ISM. Under our interpretation of the X-ray data and infrared data highlighting different components of the NSR, this outer shell would form the boundary of the infrared cavity. The density of the shell for all of the runs is moderately higher than the density of the surrounding medium ($6\times$ for $n=0.001$, $2\times$ for $n=0.01$, $4\times$ for $n=0.1$, $40\times$ for $n=1$ and $9\times$ for $n=10$), but is extremely greater than the density of the cavity (e.g. a factor of $2.4\times10^4$ for $n=0.001$ and $3.1\times10^7$ for $n=10$) and pile-up regions (a factor of $4.6\times10^3$ for $n=0.001$ and $6.8\times10^6$ for $n=10$). Small spikes of higher pressure are seen in the highest density regions of the outer shell (see top right panel of Figure~\ref{V745 Sco NSR simulation}) and also coincide with relatively low temperatures across the inner region of the outer shell (${\sim}9.4 \times 10^3$\,K for $n=0.001$ and ${\sim}10$\,K for $n=10$). Very low velocities are seen within the shell for all runs with the younger NSR (Run 1) exhibiting velocities of ${\sim}20 \ \text{km} \ \text{s}^{-1}$ compared to the older NSR (Run 5) with velocities of ${\sim}1 \ \text{km} \ \text{s}^{-1}$.

\subsection{X-ray modelling}\label{X-ray modelling}
As indicated, we would expect considerable X-ray emission from the NSR, just as detected in Figure~\ref{V745 Sco NSR eROSITA}, due to the high levels of shock-heating within the ejecta pile-up regions. Therefore, as in previous modelling studies of NSRs \citep{2019Natur.565..460D,2023MNRAS.521.3004H,2024MNRAS.529..236H}, we model such X-ray emission using \texttt{XSPEC} to compare with the X-ray luminosity upper limit from the eROSITA observation (see Section~\ref{X-ray flux upper limit of the nova super-remnant}). We chose to model the same energy bands as for eROSITA, namely $0.2-0.5$ keV, $0.5-1.0$ keV and $1.0-2.3$ keV. The synthetic X-ray luminosity radial profiles summing the three bands are provided in Figure~\ref{Synthetic x-ray profile}.

As seen in Figure~\ref{Synthetic x-ray profile}, the radial distribution of simulated X-ray emission for the modelled NSR in each ISM density shows that the centres of the structures have high luminosities as a result of the nova eruptions emanating from the central binary. This high X-ray luminosity in each model then gives way to a dramatic drop to almost negligible levels as the ejecta freely expand within the cavity regions in which little resistance is experienced. At the respective ejecta pile-up boundaries of each model, the X-ray emission significantly reestablishes ($4.3\times10^{21} \ \text{erg} \ \text{s}^{-1}$ for $n=0.001$ to $5.6\times10^{22} \ \text{erg} \ \text{s}^{-1}$ for $n=10$) as the new ejecta finally collide with the decelerated debris of thousands of prior eruptions and shock-heat the higher density boundary. The X-ray luminosity then continues to increase by up to $100\times$ (for the $n=10$ model) towards the NSR shells as a growing number of ejecta interact and shock-heat the pile-up region. As the density and temperatures are relatively constant across the whole pile-up region (see Figure~\ref{V745 Sco NSR simulation}), directly resulting from the continual supply of new ejecta colliding and shocking present material, bright X-ray emission radiating from across this region is feasible.

At the inner edges of the modelled NSRs, the X-ray luminosities then peak, as shown in all panels of Figure~\ref{Synthetic x-ray profile}, ranging from $\text{L}_{\text{X-ray}} = 2.7\times10^{26} \ \text{erg} \ \text{s}^{-1}$ for $n=0.001$ to $\text{L}_{\text{X-ray}} = 2.9\times10^{28} \ \text{erg} \ \text{s}^{-1}$ for $n=10$. The total X-ray luminosities predicted from such NSRs are $6.1\times10^{28} \ \text{erg} \ \text{s}^{-1}$, $4.0\times10^{28} \ \text{erg} \ \text{s}^{-1}$, $5.2\times10^{28} \ \text{erg} \ \text{s}^{-1}$, $8.5\times10^{28} \ \text{erg} \ \text{s}^{-1}$ and $1.7\times10^{29} \ \text{erg} \ \text{s}^{-1}$ for $n=0.001, 0.01, 0.1, 1$ and 10, respectively.

Evidently, the brightest NSR model (grown within the $n=10$ environment) is considerably fainter than both the observed X-ray luminosity of the shell around V745 Sco (${\sim}1.6\times 10^{34}$\,erg\,s$^{-1}$) and the X-ray luminosity predictions from previous modelling of the M31N 2008-12a NSR (${\sim}3\times10^{29} \ \text{erg} \ \text{s}^{-1}$ in \citealt{2019Natur.565..460D} and ${\sim}1\times10^{31} \ \text{erg} \ \text{s}^{-1}$ in \citealt{2023MNRAS.521.3004H}) and the KT Eri NSR \citep[${\sim}6.5\times10^{32} \ \text{erg} \ \text{s}^{-1}$;][]{2024MNRAS.529..236H}.

The five-order-of-magnitude discrepancy between the predicted and observed X-ray luminosities is large but not surprising given the fundamental limitations of one-dimensional smooth-shell models. Two physical effects, absent from the simulations, are expected to substantially boost X-ray emission. First, Rayleigh-Taylor and Kelvin-Helmholtz instabilities at the ejecta-shell interface fragment the swept-up ISM into dense clumps with a volume filling factor $f{\ll}1$ \citep{1983ApJ...274..152V,1995ApJ...444..312C}. Since thermal bremsstrahlung luminosity scales as $n^2V$, concentrating emission into clumps of density $n$(clump) $\sim$ $n$(mean)/$f$ while reducing their volume by $f$ yields a net luminosity boost of ${\sim}1/f$; for $f{\sim}10^{-2}$, this alone gives a factor of ${\sim}100$. The highly resolved shell of GK Per \citep{2012AJ....143..143S} makes this $f$ very reasonable. Second, radiative shock compression at the ejecta pile-up boundary can locally compress post-shock gas by factors of $\chi\sim10^2-10^3$ beyond the standard Rankine-Hugoniot factor of 4 \citep{1998ApJ...500..342B}, and since L$_X\sim n^2V \sim \chi n_0^2V_0$, this contributes another factor of $10^2-10^3$ to the emission. Combining these two effects multiplicatively yields a plausible total boost of $10^4-10^5$, which approximately matches the observed $10^5$ discrepancy.

The predicted NSR X-ray luminosities from our one-dimensional simulations therefore represent firm lower bounds rather than realistic estimates. The gap between model and observational upper limit should be interpreted not as evidence against the NSR interpretation, but as a direct consequence of smooth-shell, spherically symmetric hydrodynamics failing to capture the clumping and radiative compression physics that dominate real NSR X-ray emission. Constraining the true X-ray luminosity of NSRs will ultimately require either 3D hydrodynamic simulations (see e.g. the modelled single eruption of the nova V407 Cygni in \citealt{2012MNRAS.419.2329O} as well as the X-ray emission predicted from simulations of RN eruptions from RS Oph in \citealt{2026A&A...714A..23M}), or a confirmed X-ray detection of a known NSR from which empirical calibration is possible -- precisely what this detection of the V745 Sco NSR candidate would, if confirmed, begin to provide.
\begin{figure}
\centering
\includegraphics[width=\columnwidth]{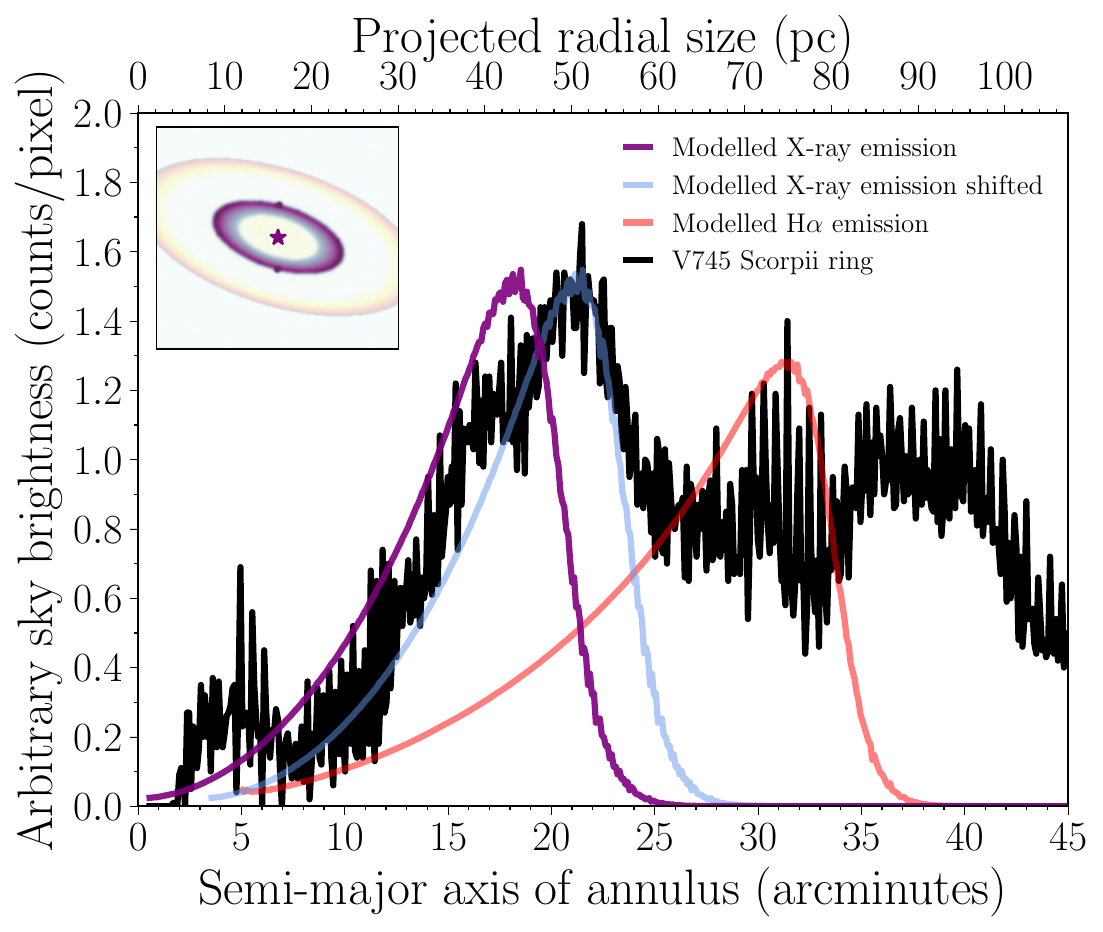}
\caption{Scaled radial profile of synthetic X-ray ring (purple line) alongside the emission radial profile (black line) of the X-ray ring around V745 Sco (from Figure~\ref{X-ray radial profile}). The modelled radial profile appears to match the profile of the observed X-ray ring, which is emphasised by shifting the radial profile of the synthetic X-ray ring (blue line). The radial profile derived from the synthetic H$\alpha$ shell is shown with a red line. The inset shows the synthetic X-ray ring (purple) generated from the $n=10$ simulation X-ray emission (shown in the bottom middle panel of Figure~\ref{Synthetic x-ray profile}) alongside the surrounding synthetic H$\alpha$ shell (orange) also derived from the $n=10$ simulation. The bright spots on the synthetic X-ray ring (at $0^{\circ}$ and $180^{\circ}$) in the inset are artifacts generated when converting the 1D radial profile to a projected 2D image.}
\label{Modelled X-ray radial profile}
\end{figure}

To preliminarily explore whether an NSR with the same structure as those grown with hydrodynamic modelling could be responsible for an X-ray ring, we first interpolated the predicted one-dimensional X-ray emission radial profile from Run 5 ($n=10$), as it is the brightest and most extended (along with Run 4). This profile was then mapped onto a two dimensional grid by computing an elliptical radius to match the elliptical geometry and position of the observed X-ray ring and assigning surface brightness values from the interpolated profile. As it is feasible for bright X-ray emission to be emanating from across this region, we chose to map the modelled X-ray emission from the NSR during the epoch when its radius was 50 pc. This elliptical mapping was smoothed to match the eROSITA image resolution, thereby creating a synthetic `on-the-sky' image (see the purple ring in the inset of Figure~\ref{Modelled X-ray radial profile}). Then, as in Section~\ref{Annular photometry of the nova super-remnant}, annular photometry was performed on the synthetic image to obtain a comparative radial profile. The resulting radial profile of this synthetic X-ray ring is provided in Figure~\ref{Modelled X-ray radial profile} (purple line) alongside the radial profile of the X-ray ring from the eROSITA data obtained through annular photometry (Figure~\ref{X-ray radial profile}). We also provide the synthetic radial profile shifted (blue line) to overlap with the radial profile of the X-ray ring to emphasise the similarities of the shape.

By inspection, it is clear that the radial profile of the synthetic ring has a slope from the cavity to the peak emission with a similar gradient to the radial profile of the observed ring. Likewise, the profile beyond the peak mimics the observed data, until reaching the surrounding ISM where the isolated modelled ring drops to zero emission. With such matching characteristics, it is feasible that a structure with the morphology of an NSR modelled in Section~\ref{Results of simulations} could be responsible for such a ring-like structure.

Additionally, using this same method, we explored whether an NSR grown with hydrodynamic modelling in Run 5 could also be responsible for the outer boundary of the larger infrared cavity, by using H$\alpha$ emission as a proxy. To predict the H$\alpha$ flux, we followed the methodology detailed in \cite{2023MNRAS.521.3004H} and \cite{2024MNRAS.529..236H}, with the H$\alpha$ extinction ($A_{\text{H}_{\alpha}} = 1.749$) applied towards V745 Sco and a distance of 8.2\,kpc. The total predicted H$\alpha$ luminosity from the radial distribution of this simulated NSR is $L_{\text{H$\alpha$}} \simeq 1.0 \times 10^{33}$ erg s$^{-1}$. Then, using the same technique as used with the synthetic X-ray emission above, we generated a synthetic `on-the-sky' image (shown in the inset of Figure~\ref{Modelled X-ray radial profile} as an orange ring) for the NSR H$\alpha$ emission and performed the same annular photometry as before to obtain a radial profile. As a first-order approximation, we find that the configuration of the synthetic X-ray ring with emission peaking at ${\sim}50$\,pc surrounded by the outer synthetic H$\alpha$ ring at ${\sim}80$\,pc, used as a proxy for the outer boundary of the infrared cavity, matches the observations shown in Figure~\ref{V745 Sco NSR eROSITA} and the IRIS $60 \ \mu $m panel of Figure~\ref{IR images}.

\begin{figure*}
\centering
\includegraphics[width=0.8\textwidth]{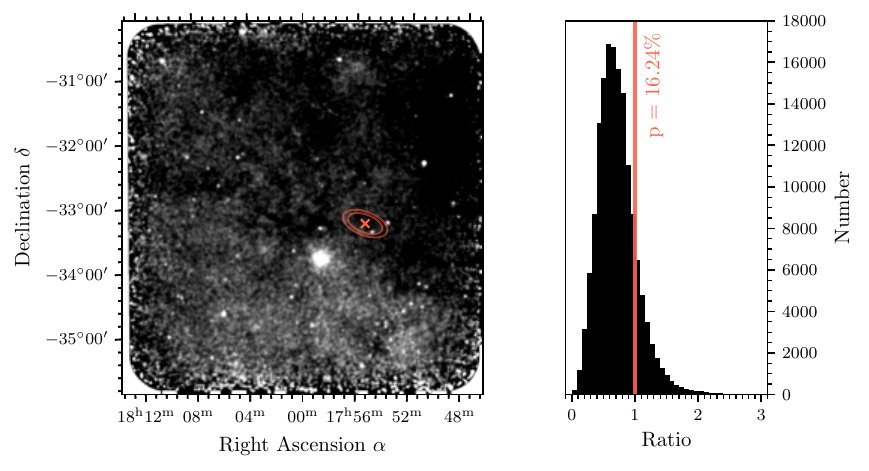}
\caption{Left: $0.5 - 1.2$ keV $5.6^{\circ} \times 5.8^{\circ}$ field of view eROSITA image (one component of image in Figure~\ref{V745 Sco NSR eROSITA}) with the location of V745\,Sco given by the red cross. The best-fitting annulus $\mathscr{E}_{\text{annulus}}$ (see text) is illustrated by the red annulus. Right: Histogram illustrating the distribution of annulus ratios $\mathscr{R}$ computed at 155,043 locations within the eROSITA data (left). The horizontal red line denotes $\mathscr{E}_{\text{annulus}}$.}
\label{Coincidence}
\end{figure*}

\section{Determining association between X-ray structure and V745 Scorpii}\label{Determining association between X-ray structure and V745 Scorpii}
As the eROSITA image covers a large area of the sky ($5.6^{\circ} \times 5.8^{\circ}$), there are many regions in the data that appear to contain structures. For this reason, we must explore a simple coincidental alignment between V745 Sco and the X-ray structure centred on its location. As such, we follow the same method employed in \citet{2024MNRAS.529L.175H}, used to determine the significance of a potential far-infrared NSR cavity being found at the location of the RN RS Ophiuchi.

We firstly define a region which will be used as a reference to compare with other regions of the eROSITA image. As the X-ray structure has a ring-like shape, we chose an annulus as an appropriate region. We now introduce a ratio, $\mathscr{R}$, defined as the ratio of the flux per unit sky area within the annulus to the flux per unit sky area in the surrounding and interior (cavity) regions. For this analysis, we masked all sources above a threshold to remove spurious contributions being made to the ratio (e.g. the two bright sources close to the location of V745 Sco).

For the reference annulus, we fixed the position to the location of V745 Sco and fixed the ellipticity as 0.5 to match the observed shell (semi-major and semi-minor axes of 19 arcmins and 10 arcmins, respectively). The semi-major axis was allowed to vary between $5 - 30$ arcmin with 1 arcmin increments and position angle was allowed to range between $0 - 180^{\circ}$ with $1^{\circ}$ increments. The width of the annulus was fixed at 30\% of the annulus semi-major axis and the width of the surrounding ellipse was fixed to 5$\times$ the annulus semi-major axis.

We find that a reference annulus fixed to the position of V745 Sco produces the largest ratio of $\mathscr{R} = 0.0262$ with a semi-major axis of 17 arcmins and a position angle of $159^{\circ}$. In Figure~\ref{Coincidence}, we illustrate this best fitting annulus and refer to it as $\mathscr{E}_{\text{annulus}}$. This annulus defines our shell around V745 Sco as it contains the majority of emission from the ring, though we note the lower segment of the shell is not fully accounted for.

For the Monte Carlo analysis, we sampled $\sim$155,000 annuli on a uniformly distributed spatial grid across an inner $4.1^{\circ} \times 4.7^{\circ}$ region of the eROSITA image (as the analysis was constrained at the outer edges of the image by the width of the surrounding ellipse), each with a randomly selected ellipticity, position angle and semi-major axis, and computed $\mathscr{R}$ at each position. The range of ellipticities is derived from the structure having an inclination between $0-70^{\circ}$; the range of position angles is between $0-180^{\circ}$, and the semi-major axis is taken from a range of radial sizes between $23.85 - 71.56$ pc ($10 - 30$ arcmins). The distribution of the computed $\mathscr{R}$ for all $\sim$155,000 annuli is shown in the right panel of Figure~\ref{Coincidence} and is normalised to the ratio $\mathscr{R} = 0.0262$.

We find that 83.76\% of the sample of annuli (without any fixed parameters) have a ratio $\mathscr{R} < 0.0262$. As such, the probability of a chance alignment of V745 Sco with a structure that has more X-ray emission than the $\mathscr{E}_{\text{annulus}}$ ring is $16.24\%$, with a large proportion (${\sim}$38\%) of these annuli clustered around the significantly bright source to the southeast of the nova's location.

The 16.24\% chance alignment probability for the X-ray ring (which we denote as $P(X)=0.1624$) alone should not be interpreted in isolation. Each of the additional lines of evidence presented in this work carries its own physical constraints, and their conjunction is far more constraining than any single test. The probability of finding an infrared cavity with a matching position angle (within $\pm5^{\circ}$) spatially coincident with the X-ray shell is $\sim$4.2\% (denoted as $P(I)=0.042$; Section~\ref{Probability of finding coincident extended object in IR data}). Further, the radial profile of the X-ray ring closely tracking that of the confirmed M31N 2008-12a NSR in both shape and scale (Section~\ref{Annular photometry of the nova super-remnant}) is unlikely to arise from a noise fluctuation or unrelated extended source, to which we conservatively assign a probability of 15\%, interpreted as the conditional probability $P(RP\,|\,X)=0.15$.

We now evaluate the joint probability of these combined constraints by considering two limiting scenarios. In these scenarios, we also incorporate a geometric alignment constraint, where the centre of a random X-ray structure must fall within a conservative radius of $5^{\prime}$ from V745 Sco. Given the $4.1^{\circ} \times 4.7^{\circ}$ field, this corresponds to an alignment probability of $0.113\%$ (from $78.54 \text{ arcmin}^2 / \, 69,372 \text{ arcmin}^2$), which we denote as $P(A)=0.00113$. In the first scenario, we assume all properties are fully correlated and therefore, to avoid compounding their probabilities, we adopt the most conservative constraint, given by the X-ray ring chance alignment (16.24\%). This yields a joint probability of $P \approx P(X) \, P(A) = 0.1624 \times 0.00113 \approx 1.84 \times 10^{-4}$. In the second scenario, we treat the radial profile as conditional on the presence of an X-ray structure, i.e. $P(RP\,|\,X)$, while treating the infrared cavity and geometric alignment as independent constraints. As such, the joint probability becomes $P = P(X) \, P(RP\,|\,X) \, P(I) \, P(A) = 0.1624 \times 0.15 \times 0.042 \times 0.00113 \approx 1.16 \times 10^{-6}$.

While the individual assumptions in such calculations are uncertain, the conclusion is robust across both scenarios: the convergence of spatial coincidence, matching position angle, and NSR-like radial morphology is far more difficult to attribute to chance than the 16.24\% figure alone implies. While the resulting probabilities should be taken as order-of-magnitude estimates rather than exact values, they are consistently indicating that a chance alignment is unlikely and thus make a compelling case in favour of a genuine NSR detection.

\section{Other origins of structure}\label{Other origins of structure}
If the structure is indeed real then we must also rule out other possible astrophysical phenomena with similar characteristic sizes as NSRs, including supernova remnants (SNRs) and superbubbles, as well as other responsible sources.

We would expect strong radio emission if the structure were a SNR due to synchrotron radiation from the outer shock front \citep[high velocity shocks at 1000\,s km\,s$^{-1}$ combined with strong magnetic fields, e.g.][]{2017hsn..book.2041D}, yet NSRs will have a weaker magnetic field and exhibit outer shell expansion velocities of ${\sim}10$\,km\,s$^{-1}$, not energetic enough to accelerate electrons to relativistic velocities and produce strong radio emission. As such, the X-ray ring's location was examined in various radio surveys. The structure's location was observed neither with MeerKAT array \citep{2016mks..confE...1J} nor in the ASKAP Evolutionary Map of the Universe survey \citep[ASKAP EMU;][]{2011PASA...28..215N,2021PASA...38...46N}. As noted in Section~\ref{Cavity structure in infrared data}, its location was covered by the Rapid ASKAP Continuum Survey \citep[RACS;][]{2020PASA...37...48M,2021PASA...38...58H} but within these data, there was no radio emission detected coincident with the X-ray ring. As such, we can strongly conclude that we are not detecting a faint SNR in the eROSITA data. 

\begin{figure}
\centering
\includegraphics[width=0.9\columnwidth]{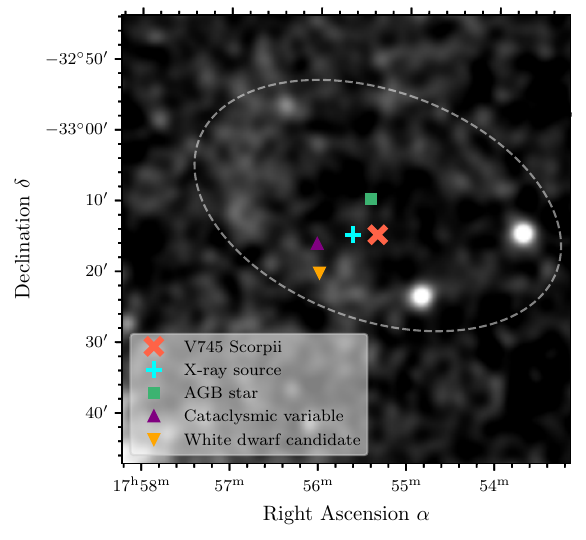}
\caption{Sources close to V745 Sco within a 10 arcminute radius taken from the SIMBAD database (as described in the text).}
\label{Nearby sources}
\end{figure}

We also cross-matched the location of the candidate X-ray NSR with both the most recent sample of 310 Galactic SNRs from \citet{2025JApA...46...14G} and 132 SNRs given in the Chandra Supernova Remnant Catalog\footnote[3]{https://hea-www.cfa.harvard.edu/ChandraSNR/snrcat gal.html} and found that none of the remnants overlapped with the X-ray structure as all SNRs were confined to the Galactic plane. According to \citet{2022MNRAS.514..728A}, the X-ray luminosities of many Galactic, LMC and SMC SNRs\footnote[4]{See Figure 2 of \citealt{2022MNRAS.514..728A}.} are much higher ($10^{35} - 10^{37}$\,erg\,s$^{-1}$) than both the luminosity of the X-ray ring (${\sim}1.6\times 10^{34}$\,erg\,s$^{-1}$) and the total emission within the region (${\sim}2.3\times 10^{34}$\,erg\,s$^{-1}$; see Section~\ref{Annular photometry of the nova super-remnant}). Yet, a good proportion have X-ray luminosities of a comparable (e.g. SN\,1006 with ${\sim}3.4\times 10^{34}$\,erg\,s$^{-1}$; \citealt{2013ApJ...771...56U,2022MNRAS.514..728A}, G85.9-0.6 with ${\sim}1-4\times 10^{34}$\,erg\,s$^{-1}$; \citealt{2008ApJ...674..936J}, and MCSNR J0524–6624 with ${\sim}1.3\times 10^{34}$\,erg\,s$^{-1}$ \citealt{2024A&A...692A.237Z}), or even lower (e.g. G38.7-1.4 with ${\sim}7.29\times 10^{33}$\,erg\,s$^{-1}$; \citealt{2014ApJ...785..118H,2022MNRAS.514..728A}) magnitude. Taking the values above as representative of the whole sample of SNR X-ray luminosities, it appears that the Galactic, LMC and SMC SNR X-ray luminosity distribution firmly extends into the regime of the candidate structure, therefore its X-ray luminosity alone is not enough to rule out a faint SNR origin, but rather its lack of radio emission is by far the stronger discriminant.

We explored the possibility of the structure being a superbubble by inspecting the locations of the most well-studied OB associations from \citet{2020NewAR..9001549W}. We found that zero OB associations are coincident with the location of the X-ray shell and therefore, along with the non-detection of radio emission, rule out that it is a superbubble formed by an OB association. 

The SIMBAD database \citep{2000A&AS..143....9W} was utilised to search a radius of 10 arcminutes around V745 Sco to find any other sources potentially responsible for the X-ray ring. Aside from the multitude of variable stars and eclipsing binaries within the ${\sim}1200$ sources, there were a handful of notable objects (see Figure~\ref{Nearby sources}) including the X-ray source NuSTAR J175538-3314.9 \citep{2017ApJ...836...99L}; the cataclysmic variable OGLE BLG-DN-306 G357.4-04.0-44 \citep{2015AcA....65..313M}; a white dwarf candidate \citep{2021MNRAS.508.3877G}; and an asymptotic giant branch (AGB) star \citep{2020ApJ...892...52L}. 

While AGB stars do lose material via dusty winds that go on to create extended envelopes \citep{2012A&A...537A..35C} and bubbles \citep{2023MNRAS.522..811O} within the surrounding ISM, these structures are predominantly less than one parsec across \citep{2012A&A...537A..35C}. Secondly, if the X-ray source is an X-ray binary, then it may also be surrounded by an extended shell as a result of winds and outbursts \citep[as is the case for the 9.5$\times$14.3\,pc shell around CI Cameleopardalis;][]{Fesen_2024} or from interaction of strong winds with ISM generating bow shocks \citep{2018MNRAS.474.4421G}, however it is not known if NuSTAR J175538-3314.9 is an X-ray binary. Cataclysmic variables can indeed host shells if they are systems exhibiting dwarf nova \citep[e.g.][]{2007Natur.446..159S,2012ApJ...758..121S,2024MNRAS.529..212S} or nova eruptions \citep[e.g.][]{2025MNRAS.539..246S} but these structures are of sub-parsec scales. Lastly, the white dwarf candidate may be surrounded by a planetary nebula, and though they do emit in the X-ray regime as a result of self-shocking winds \citep{2012AJ....144...58K}, again the expected sizes of such nebulae are much less than one parsec \citep{2003A&A...405..627T}, much smaller than the X-ray shell.

Therefore, it is unlikely that any of these objects in the vicinity could be responsible for the creation of a much larger ${\sim}90 \times 50$\,pc X-ray ring within a ${\sim}167 \times 120$ pc outer shell; if a real structure, the most likely source responsible is the RN, V745 Sco, which has the capability to create a nova super-remnant of this scale via repeated nova eruptions, and in the process, generating X-ray emission through shock-heating.

\section{Conclusions}\label{Conclusions}
In this paper, we present the identification of a shell in eROSITA X-ray imaging around the Galactic recurrent nova V745 Scorpii, that we propose to be another Galactic nova super-remnant, and the first to be initially captured through its X-ray emission. The indicators of the structure being a nova super-remnant detailed in this work include:

\begin{enumerate}[label=(\roman*), leftmargin=*]
\item it is aligned with V745 Sco, and the nova is in a central location with respect to the structure;
\item its size is consistent with the scales of other NSRs in the Galaxy, M31 and the LMC;
\item the brightness radial profile of the X-ray ring is reminiscent of the H$\alpha$ brightness radial profile of the prototype NSR in M31 thus indicating a common morphology;
\item there is a larger cavity structure in infrared imaging spatially coincident with, and potentially related to, the X-ray NSR with a similar position angle;
\item and a synthetic `on-the-sky' NSR derived from the predicted X-ray emission of a modelled V745 Sco NSR mimics the observed X-ray ring.
\end{enumerate}

If this structure is indeed a nova super-remnant belonging to V745 Sco, then much deeper follow up in the optical (and other wavebands) should better reveal the structure, with dimensions of ${\sim}167 \times 120$\,pc matching the outer infrared shell, and support our finding of the X-ray emission from the nova super-remnants's inner component within eROSITA data.

\begin{acknowledgements}
We thank the anonymous reviewer for their insightful comments. This work is based on data from eROSITA, the soft X-ray instrument aboard {\it Spektrum-Roentgen-Gamma}, a joint Russian-German science mission supported by the Russian Space Agency (Roskosmos), in the interests of the Russian Academy of Sciences represented by its Space Research Institute (IKI), and the Deutsches Zentrum für Luft- und Raumfahrt (DLR). The SRG spacecraft was built by Lavochkin Association (NPOL) and its subcontractors, and is operated by NPOL with support from the Max Planck Institute for Extraterrestrial Physics (MPE). The development and construction of the eROSITA X-ray instrument was led by MPE, with contributions from the Dr. Karl Remeis Observatory Bamberg \& ECAP (FAU Erlangen-Nuernberg), the University of Hamburg Observatory, the Leibniz Institute for Astrophysics Potsdam (AIP), and the Institute for Astronomy and Astrophysics of the University of Tübingen, with the support of DLR and the Max Planck Society. The Argelander Institute for Astronomy of the University of Bonn and the Ludwig Maximilians Universität Munich also participated in the science preparation for eROSITA. The eROSITA data shown here were processed using the eSASS software system developed by the German eROSITA consortium. This scientific work uses data obtained from Inyarrimanha Ilgari Bundara / the Murchison Radio-astronomy Observatory. We acknowledge the Wajarri Yamaji People as the Traditional Owners and native title holders of the Observatory site. CSIRO’s ASKAP radio telescope is part of the Australia Telescope National Facility (https://ror.org/05qajvd42). Operation of ASKAP is funded by the Australian Government with support from the National Collaborative Research Infrastructure Strategy. ASKAP uses the resources of the Pawsey Supercomputing Research Centre. Establishment of ASKAP, Inyarrimanha Ilgari Bundara, the CSIRO Murchison Radio-astronomy Observatory and the Pawsey Supercomputing Research Centre are initiatives of the Australian Government, with support from the Government of Western Australia and the Science and Industry Endowment Fund. This paper includes archived data obtained through the CSIRO ASKAP Science Data Archive, CASDA (https://data.csiro.au). MS and SS acknowledges support from the Deutsche Forschungsgemeinschaft through the grants SA 2131/13-1, SA 2131/14-1, and SA 2131/15-1. MMS acknowledges the support of United States National Science Foundation Grants 1910001, 2107954, 2108234, 2407763, and 2407764. This study made use of Prospero high-performance computing facility at Liverpool John Moores University. This research has made use of the SIMBAD database, operated at CDS, Strasbourg, France \citep[see][]{2000A&AS..143....9W}. This research has made use of the VizieR catalogue access tool, CDS, Strasbourg, France. The original description of the VizieR service was published in \citet{2000A&AS..143...23O}.
\end{acknowledgements}

\bibliographystyle{aa}
\bibliography{bibliography.bib}

\begin{appendix}
\section{V745 Sco NSR evolutionary timescale}\label{V745 Sco NSR evolutionary timescale}
From the modelling of the NSR around V745 Sco, detailed in Section~\ref{Modelling of V745 Scorpii NSR}, we find that in an ISM with a density of $n=10$ that it would take ${\sim}6.4\times10^{7}$\,yr to grow the NSR shell to the observed size. Here, we briefly consider whether this timescale is feasible considering the mass of the system and whether the donor would be able to maintain its mass loss for this period.

Regarding the amount of mass available to the accreting white dwarf over ${\sim}6.4\times10^{7}$\,yr: assuming a $1\,\text{M}_{\odot}$ WD, then we would require $0.4\,\text{M}_{\odot}$ of retained material to reach the Chandrasekhar limit (the point at which the nova eruptions would cease, and the NSR shell would not have new ejecta to drive its growth). If we also assume a mass accumulation efficiency of ${\sim}50\%$, then we would require $0.8\,\text{M}_{\odot}$ of material lost by the donor. A recent estimate of the quiescent accretion rate within the V745 Sco system was found to be $5\times10^{-8}\,\text{M}_{\odot}\,\text{yr}^{-1}$ \citep{2026arXiv260901388J}. For $0.8\,\text{M}_{\odot}$ of matter to be accreted at this rate, the system would require ${\sim}1.6\times10^{7}$\,yr $-$ of the same order of magnitude as the $n=10$ simulation. Furthermore, we note that the mass accretion rate of nova systems will vary over their long-term evolution and this may allow for a more extended period of mass accretion.

Regarding whether the companion star within the V745 Sco system can maintain mass loss for ${\sim}6.4\times10^{7}$\,yr, we note that while this timescale may be longer than the typical red giant branch lifetime for higher mass stars, mass loss within the binary system would alter the donor’s evolution and thereby increase the donor’s time on the red giant branch. However, the precise extent of this evolutionary lengthening remains an uncertainty.

\section{Radiative cooling energy losses}\label{Radiative cooling energy losses}
To determine the amount of energy lost due to radiative cooling within the modelling outlined in Section~\ref{Modelling of V745 Scorpii NSR} compared to the total amount of energy injected into the simulation through mass loss, we can take the initial energy within the pre-populated ISM plus the total energy from mass loss (given in Table~\ref{Runs}) and directly compare this with the total energy within the simulation domain of the final epoch. For each of the five runs, we find the fraction of energy lost due to radiative cooling to be ${\sim}33\%$, ${\sim}27\%$, ${\sim}36\%$, ${\sim}55\%$ and ${\sim}67\%$ for ISM $n = 0.001, 0.01, 0.1, 1$ and 10, respectively.

This energy loss within the simulations due to radiative cooling would manifest as either optical emission at the NSR shell (to be searched for, as stated in Section~\ref{Conclusions}) or X-ray emission within the interior (which we detail in this current work).

\end{appendix}
\end{document}